\documentclass[conference,compsoc]{IEEEtran}

\ifCLASSOPTIONcompsoc
  \usepackage[nocompress]{cite}
\else
  \usepackage{cite}
\fi

\usepackage{amsmath}

\usepackage{array}

\usepackage[caption=false,font=footnotesize]{subfig}

\usepackage{url}
\usepackage{algorithm}
\usepackage{algorithmicx}
\usepackage{algpseudocode}
\usepackage{amsmath}
\usepackage{textcomp}
\usepackage{siunitx}
\usepackage{graphicx} 
\usepackage[numbers,compress]{natbib}
\newcommand{\opt}{\mathsf{OPT}}

\begin{document}
	\title{Modeling and Routing for Predictable Dynamic Networks
	}	
	\author{
			 \textbf{This paper is only for copyright protection, and unpublished to the top-level version.}\\
			 
			 \\

	\IEEEauthorblockN{Zengyin Yang,
		Qian Wu,
		Hewu Li, 
		Zhize Li and
		Jianping Wu,  Fellow IEEE
	}
	\IEEEauthorblockA{ Tsinghua University, Beijing, China}
	\IEEEauthorblockA{\{ yang-zy14, zz-Li14 \}@mails.tsinghua.edu.cn, \{ wuqian, lihewu, jianping \}@cernet.edu.cn}
	
}

	\maketitle

	\begin{abstract}
Currently, building a Satellite-Based Internet (SBI) providing global Internet service has become a trend, and the Iridium-like constellation is considered as the promising topology. However, such topology is dynamic in terms of node position, network connectivity and link metric, which lead to drastic end-to-end path changes in polar region and overly frequent routing updates.
In this paper, we propose a link lifetime assisted routing (LLAR) scheme that incorporates life-time of links into their metrics to solve the routing oscillation in polar region. 
However, it is unpractical to implement the routing scheme into dynamic network directly because these dynamics will require routing to be updated continuously for keeping consistent with these dynamics. 
Furthermore, in order to implement the routing scheme into such network and meanwhile decrease routing updates, we establish a topology model to formulate the dynamic topology into a series of static snapshots, and propose a dynamic programming algorithm to determine the timing of routing update. 
Extensive experiments show that our methods can significantly decrease changed paths and affected satellite nodes, greatly improving routing stability. They can also achieve better network performance, less packets loss and higher throughput
Besides, few routing updates are triggered to keep consistent with network dynamics. 
	\end{abstract}


		\section{Introduction}
	\label{Introduction}
With increases in demands of human activities and capacities of satellites, building a satellite-based Internet (SBI) providing global Internet services has become attracted lots of attentions from multiple points, including traditional space network providers~\cite{Inmarsat,www.iridium.com}, the Internet content providers~\cite{SpaceX, Facebook}, and networking equipment providers~\cite{wood2007using}. 
Among them, the Iridium system is a well-known practical network providing global service, and also it will be updated to provide IP-based services for global users~\cite{www.iridium.com}.
And the Iridium-like constellation is considered as the promising topology of SBI due to easier implementation and smaller communication delay~\cite{www.iridium.com,OneWeb, evans1998satellite}. 
In the Iridium-like constellation topology, there are two polar and satellite routers permanently fly along some fixed orbits from a polar to other.
As satellite routers are extremely far away from each other, the propagation delay of link is so large that is in most cases taken as the link metric for routing selection algorithm~\cite{akyildiz2002mlsr,jurski2009routing,ekici2001distributed,werner1997atm,henderson2000distributed}.

Actually, the Iridium-like constellation topology is dynamic.
The link metric and network connectivity continuously changes with the position of routers, which brings great challenges for SBI to achieve routing stability.
The main problems are two-fold.

1) \textit{Polar routing oscillation}. 
In polar region, link metric and link connectivity are conflicted.
That is,  when a link moves closer to the polar region, it will disconnect but its propagation delay will become smaller. And it is opposite when the link leaves the polar region. When the metrics of nearly disconnecting links and newly reconnecting links relatively vary,  many end-to-end paths will be forced to switch forth and back on these links.
In fact, the polar routing oscillation overly  occurs about once per 136.8s and 
impacts $15\% \sim 30\%$ of total paths for each oscillation, which incur extremely poor network performance, such as severe packet drops and low throughput.

2) \textit{Frequent routing updates}. 
As the topology of such network continuously changes, it will require the routing to be updated frequently for keeping consistent with the topology changes. 
If routing updates fall behind the topology changes, some path may become invalid or sub-optimized. However, in each update, at most $41\%$ of total paths will be re-routed and many limited on-board resources will be wasted. Hence, a trade-off between the timeliness of keeping consistent with topology dynamics and the cost of routing updates is needed.

These routing stability problems are essentially different from current researches in ground network or space network.
First, the polar routing oscillation is absent in the stable routing schemes specific to traditional Internet and ad-hoc network~\cite{basu2001stability, boukerche2011routing}. Previous schemes for routing stability in space network  mostly focused on the convergence issue of traditional routing protocols (e.g., BGP) suffering from topology dynamics~\cite{berson2009effect, Etefia2010Emulating}, but the issue of polar routing oscillation incurred by route selection have not been proposed yet.  
Second,	 the cost of routing update has not been considered into previous methods of maintaining the consistent with topology dynamics, like PLSR~\cite{gounder1999routing, fischer2008topology} and DT-DVTR~\cite{werner1997atm,werner1997dynamic}. Hence, the goal of minimizing routing updates and meanwhile maintaining the consistent with topology dynamics can not be achieved in these methods.
	\begin{figure*}[!t]	
	\begin{minipage}{0.31\linewidth}
		\centering
		\includegraphics[width=1\linewidth]{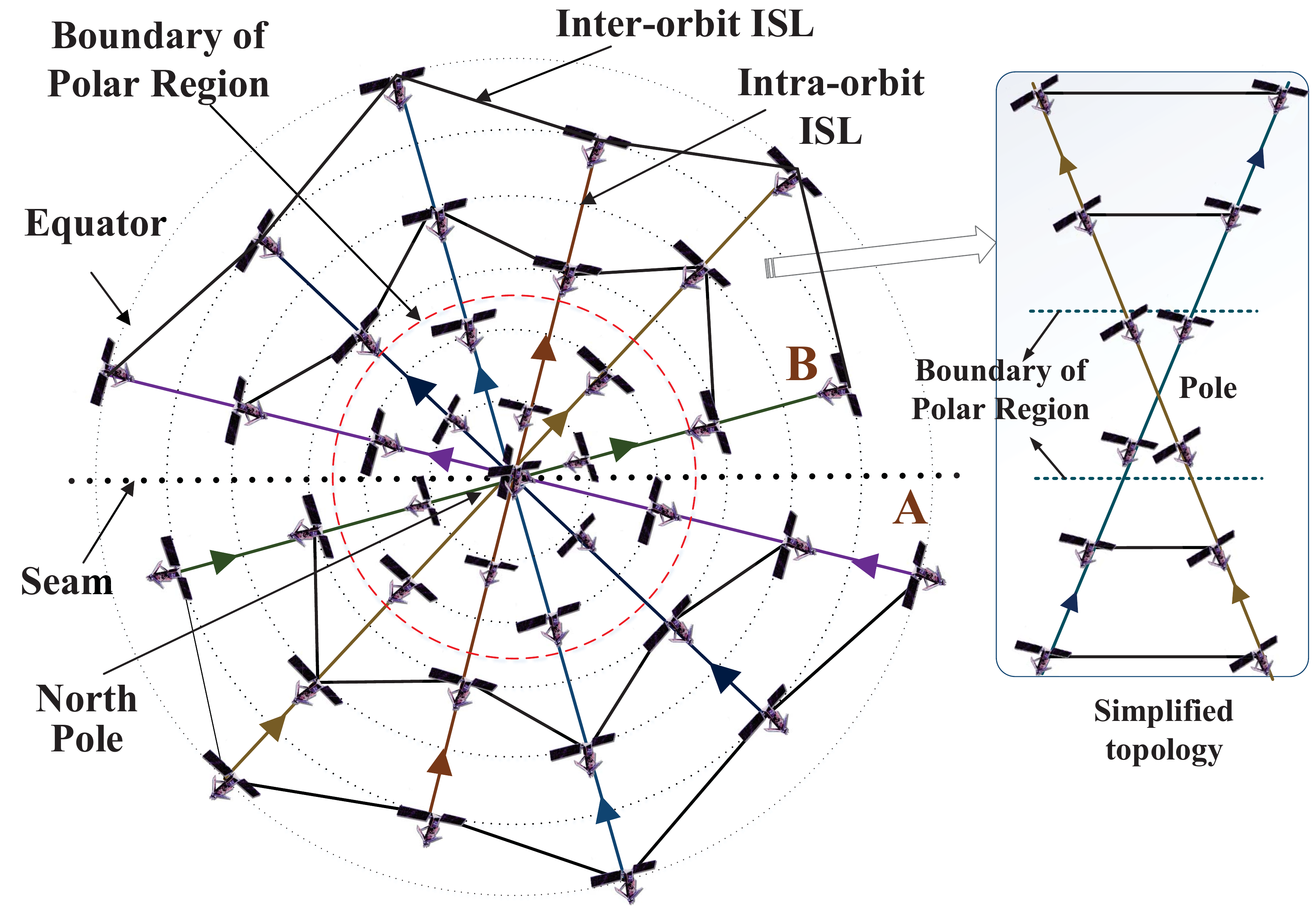}%
		\caption{Topology of Iridium-like constellation.}
		\label{Iridium_topology}
		
	\end{minipage}
	\hfill
	\begin{minipage}{0.29\linewidth}
		\centering
		\includegraphics[width=\linewidth]{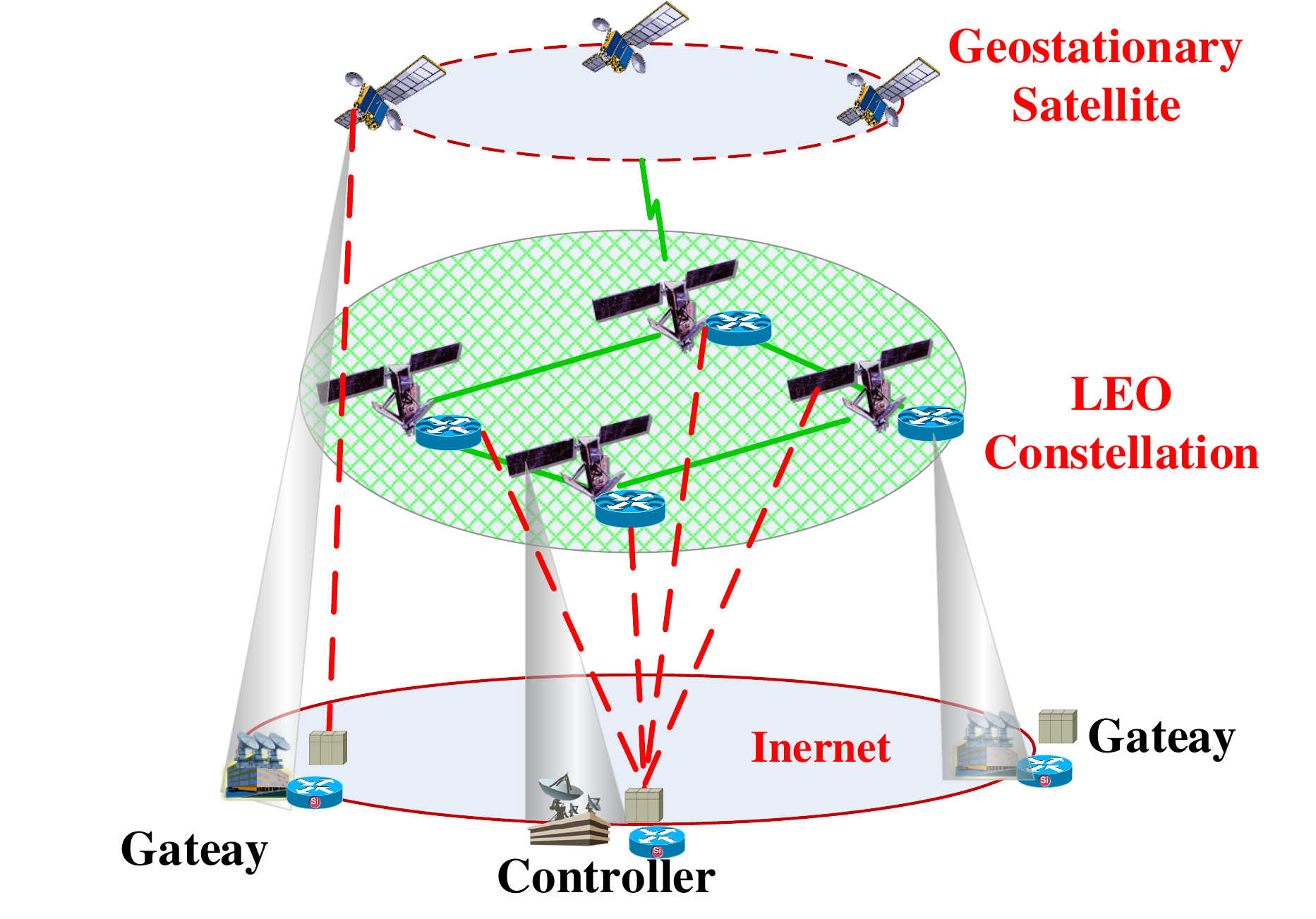}%
		\caption{Satellite-based Internet system.}
		\label{Space_Internet}
		
	\end{minipage}
	\hfill
	\begin{minipage}{0.3\linewidth}
		\centering		
		\makeatletter\def\@captype{table}\makeatother
		\caption{Parameters of selected constellation}
		\label{Parameters}
		\includegraphics[width=\linewidth]{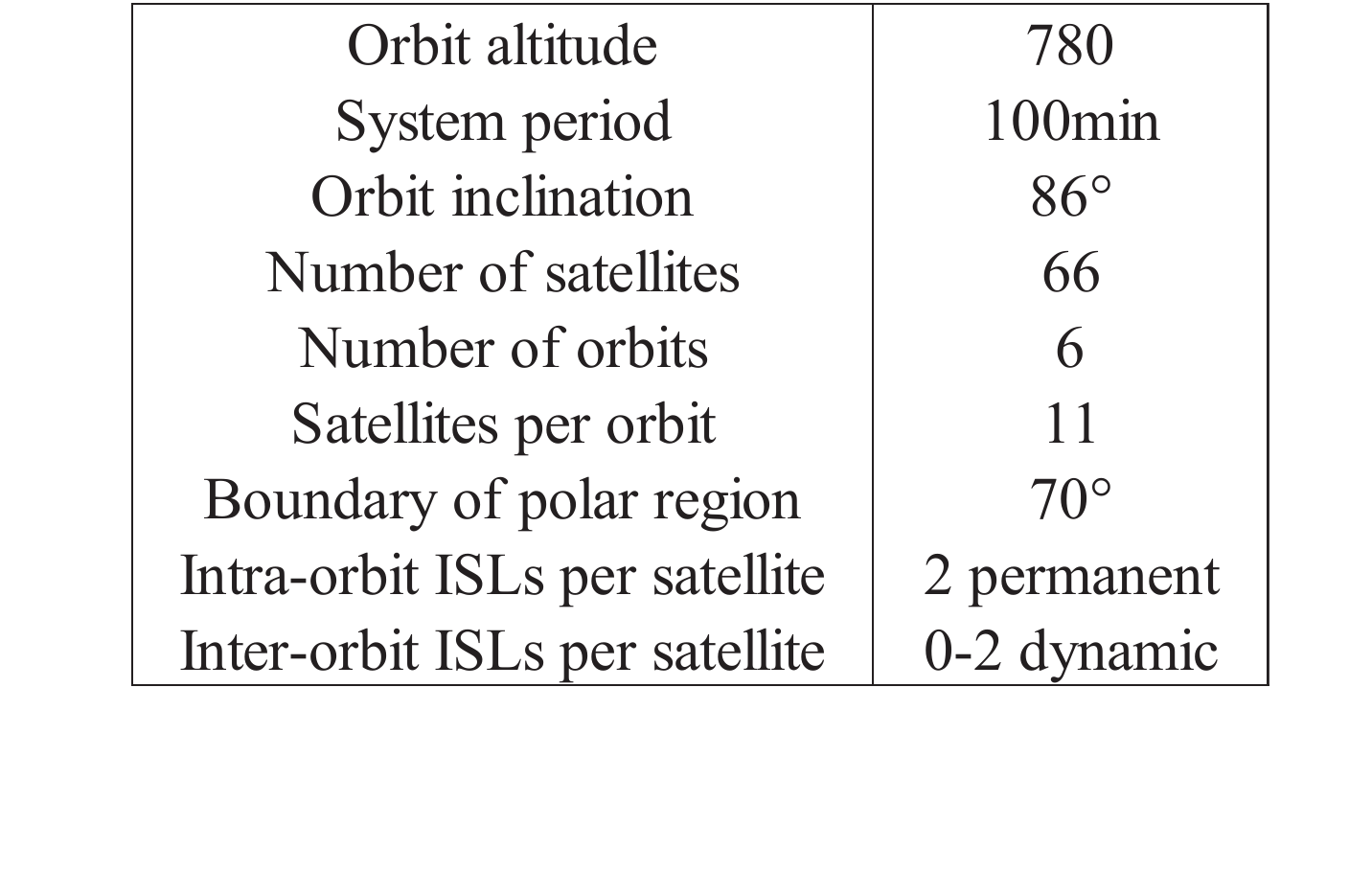}
	\end{minipage}	
\end{figure*}

In this paper, we design a \textit{link lifetime assisted routing} (LLAR) scheme to mitigate the polar routing oscillation. 
The proposed scheme firstly defines a link lifetime in LLAR to locate the links where oscillation occurs. Then it adds the lifetime of these severely dynamic links into their metrics to eliminate the variations of link metric. Furthermore, by taking advantages of the Dijkstra's greedy property, it decreases the amount of paths sharing these dynamic links and to avoid routing switching back and forth between them.
Furthermore, we adopt the Divide-and-Merge methodology \cite{cheng2005a} to avoid routing updates when the topology keeps unchanged or slightly changed. 
First, we divide the dynamic topology into sequences of static snapshots. Second, for decreasing the number of static snapshots, we propose a dynamic programming algorithm to merge those unchanged or only slightly changed contiguous static snapshots. Consequently, routing updates will only be triggered at the start time of each merged snapshot.

Extensive experiments show that our methods can significantly improve the routing stability issues in SBI.
For polar routing osculation issue, it can decrease 50\% of changed paths and affected satellite nodes than PLSR and 32\% than DT-DVTR.
That results in a decrease of more than 90\% packet loss and a increase of almost 20\% throughput compared with others.
And for routing update issue, our methods can achieve few routing updates and meanwhile keep consistent with the SBI dynamics.

The rest of this paper is organized as follows. Section \ref{Network_Topology} presents the background about iridium-like constellation and routing concept.  In section \ref{Analysis_on_Network_Stability}, the routing stability in SBI is deep analyzed. The corresponding routing scheme is presented in section \ref{Routing_for_a_Snapshot} and a detailed presentation about our topology model is provided in section \ref{Topology_Model}.
In section \ref{Experiment}, we present the experiments used to verify our methods. Finally, we review the related work in section \ref{Related_Work} and draw a conclusion in section \ref{Conclusion}.

	\section{Background}
\label{Network_Topology}

Currently, the IP-based applications of ground network and space network are integrating. Building a satellite-based Internet (SBI) providing global Internet services has become an important part of the future network. In fact, there have been some proposed systems taking IP technology as the basic operation mode, such as Inmarsat \cite{Inmarsat} and Iridium \cite{www.iridium.com}. Meanwhile, some Internet content providers actively carry out the construction of SBI, such as Google and Facebook \cite{SpaceX, Facebook}. Google has invested the companies SpaceX and OneWeb to build SBI, in order to provide Internet access in rural and remote areas \cite{SpaceX, OneWeb}. Facebook also announced the Internet.org Platform with other six companies, to bring the Internet to everybody via satellites and drones \cite{Facebook}.
Some practical systems also have been proposed \cite{www.iridium.com} and successful on-orbit Cisco router testing has been implemented \cite {wood2007using}.

\subsection{Iridium-like Constellation}
In the world of satellites, the Low Earth orbit (LEO) constellation offers significantly smaller propagation delays as well as lower signal attenuation than geostationary (GEO) satellites, therefore, the LEO constellation equipped with inter\-satellite link (ISL) has been considered as a good way to form a global communication environment in space.    
As the use of ISL greatly complicates the design of system, the LEO constellation with circular polar orbits is prone to be chosen. Up to now, most proposed LEO constellations with ISL are based on the circular polar orbits \cite{www.iridium.com, evans1998satellite, ananasso1998satellite}. Among them, the Iridium system is a only well-known practical commercial network, and also it will be updated to provide IP-based services for global users~\cite{www.iridium.com}. Thus, Iridium-like constellation is considered as a representation for analysis and simulation in this paper, whose parameters are given in TABLE \ref{Parameters}.

Fig.\ref{Iridium_topology} shows that the Iridium-like system is seen from the North Pole. The constellation has 6 orbits, all of which intersect at the pole. In each orbit, there are 6 or 7 satellites painted in the figure. However, for the overall system, there are totally 11 evenly distributed satellites in each orbit. And all satellites predictably fly along their own orbits from one polar to the other. The time of all satellite flying a circle around the earth is called the system period, about $100min$.

Each satellite is equipped with 2 to 4 ISLs. The ISL interconnects neighboring satellites within line-of-sight. Two of these ISLs (intra-orbit ISL) are used to link the ahead and behind satellite in the same orbit, while the other two (inter-orbit ISL) are used to interconnect adjacent satellites in adjacent orbits. For an inter-orbit ISL, the viewing angle for maintaining it varies with its latitude. When an inter-orbit ISL is above a certain latitude (boundary of polar region), the angular rates for tracking antennas become so high that the inter-orbit ISL will be dropped \cite{evans1998satellite}. In other words, when a satellite enters the polar region, its adjacent inter-orbit ISLs will disconnect; when it leaves the region, its adjacent inter-orbit ISLs will reconnect.  Hence, the reconnection/connection of inter-orbit ISLs will jointly change in the polar region.

The simplified topology with only two adjacent orbits is shown in Fig.\ref{Iridium_topology}. 
For simplicity, the adjacent satellites in the adjacent orbits are painted in the same horizontal direction.
We can clearly get that all of intra-orbit ISLs have the same delay, and the inter-orbit ISLs disconnect in the polar region.

As shown in Fig.\ref{Iridium_topology}, the satellites at both sides of the seam are counter-rotating. The connectivity of adjacent counter-orbiting satellites frequently changes, which puts a great stress on the design of the network. In practical systems, inter-orbit ISLs passing the seam are not yet implemented \cite{www.iridium.com, evans1998satellite}. Thus, the routing between two satellites at the both sides of the seam will go through the polar region, e.g., the routing between satellite A and B who are very close.

\textbf{Dynamics of Network Connectivity}:
 The dynamics of network connectivity incurred by ISL disconnection/reconnection are regular, which is the alternate occurrence of ISL disconnection and reconnection. The time intervals between two contiguous changes only have two types, namely the time from ISL disconnection to ISL reconnection, and the time from ISL reconnection to ISL disconnection. Interestingly, every kind of time interval only has a constant value. For the selected constellation, the time from disconnection to reconnection and the time from reconnection to disconnection, are 162.7s and 111.3s, respectively. On average, the network connectivity changes occur once per 136.8s.

\textbf{Characteristics of Link Delay}:
In the space network, any two satellites are wide apart. The propagation delay is so large that is in most cases taken as the link metric.

For the intra-orbit ISL, its propagation delay almost keeps unchanged during a system period, about 13.5ms. Meanwhile, all the intra-orbit ISLs have a nearly identical propagation delay because every satellite is evenly distributed in its own orbit.
The propagation delay of inter-orbit ISL continuously varies with ISL's latitude.  When an ISL resides over the Equator, its propagation delay becomes the largest, about 15ms.  When an ISL is closest to the polar region boundary, its delay becomes the smallest, about 9ms. However, if the inter-orbit ISL moves into the polar region, its delay becomes infinite due to the ISL disconnection.

Compared with the propagation delay the queuing delay can be directly ignored, and thus the routing computing can be independent of the queuing delay \cite {jurski2009routing}. As the routing protocol design relies on the underlying topological features of the satellite network, most current routing schemes take the propagation delay as the link metric \cite{akyildiz2002mlsr,jurski2009routing,ekici2001distributed,werner1997atm,henderson2000distributed}.

		\subsection{IP-based centralized routing concept}
	\label{centralized_routing_concept}
	Compared with the ground network, the SBI has a dynamic topology and limited on-board resources, such as computing ability, storage, and power. The routing schemes that are implemented on the satellite or rely on exchanging network information to know the real network topology will incur a heavy overhead.

	By making use of the predictability of topology, the routing tables can be computed in advance of the real network topology forming. For current Iridium system with circuit switching technology, the routing tables are computed on a ground station. When a LEO satellite flies above the station, the satellite will get its routing tables of a future period. However, we don't clear the details of their current routing technology and topology model. Moreover, the routing technology of their Iridium-Next system may be still in the stage of research. 
	
	Based on this current practical system and some studies, we suppose that the routes of SBI are also computed on the ground station.
	The LEO satellite will not only get its routing tables from the stations but also can use GEO satellites to relay its routing tables, as shown in Fig.\ref {Space_Internet}. 	The utilize of GEO can avoid implementing the stations in the remote area.
	
	The routing scheme can be implemented into SBI by taking advantage of Software Defined Network (SDN) technology~\cite{bertaux2015software}. Every satellite is a switch, and the gateway can act as the controller. 


		\section{Routing Stability Issues}
	\label{Analysis_on_Network_Stability}

	The routing stability issues in SBI are different from that of ground network~\cite{basu2001stability, boukerche2011routing}. That is, the occurrence of routing instability in SBI can be considered as a series of discrete instantaneous events because the routing can be updated immediately once the topology changes. Without the process of routing convergence, routing instability will be mainly caused by route selection.  Besides, the frequency of routing updates will be introduced to indicate the occurrence frequency of routing instability.

	In this section, we deep analyze the impacts of polar routing oscillation incurred by route selection and frequent routing updates for keeping consistent with topology dynamics.

\begin{figure}[!t]
		\centering
		\includegraphics[width=0.95\linewidth]{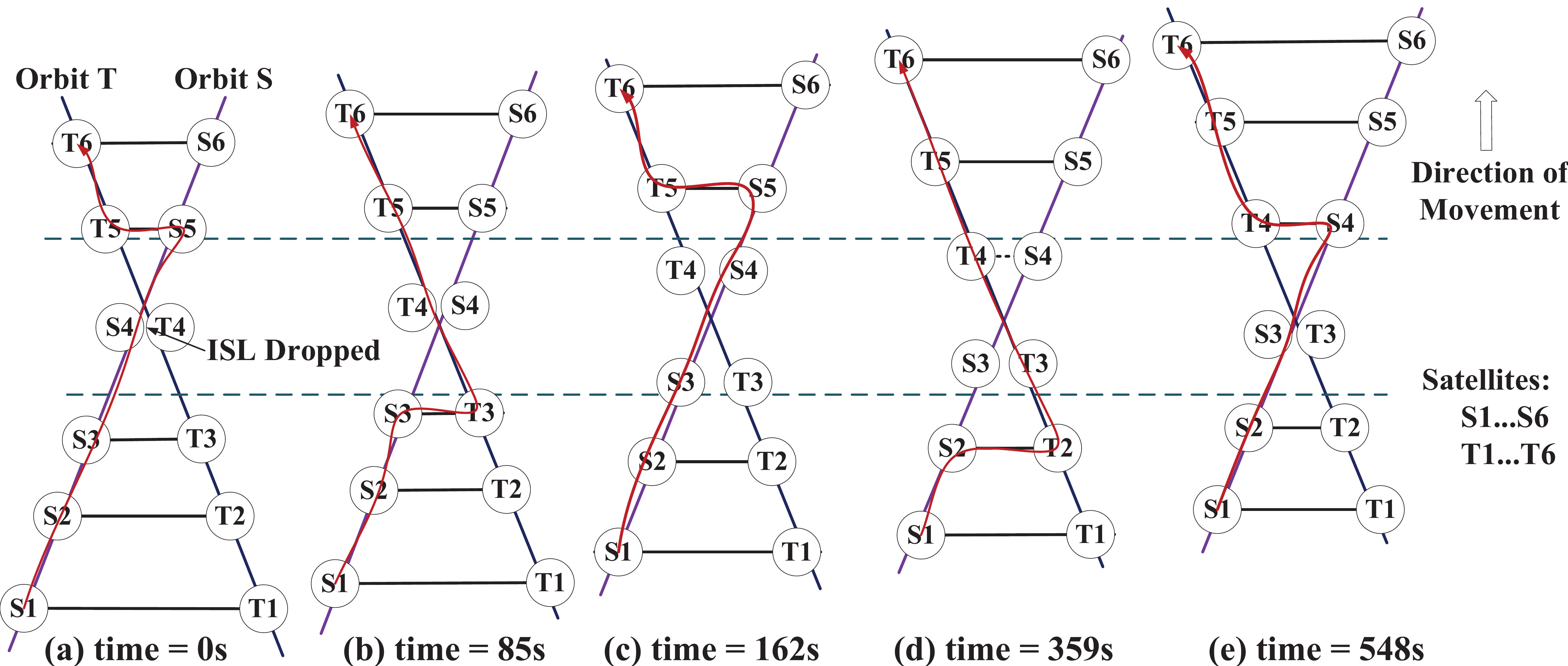}%
		\caption{Routing oscillation in the polar region}
		\label{simplified_topology_delay_change}            
	\end{figure}

	\textbf{Polar Routing Oscillation}:
	We take an example with two orbits in a polar region to analyze the impact of polar routing oscillation, as shown in Fig.\ref{simplified_topology_delay_change}. And the ISL from satellite i to j is denoted as (i, j).	
	
	For routing between two satellites in adjacent orbits, the shortest inter-orbit ISL will be chosen to adjacent orbit from the current orbit. As shown in Fig.\ref{simplified_topology_delay_change}a, the routing from satellite S1 to T6 will choose the inter-orbit ISL (S5, T5).  Since all intra-orbit ISls have the identical metric, the total sum of metrics of all chosen intra-orbit ISLs will be a constant no matter which inter-orbit ISL is chosen from ISLs (S1, T1), (S2, T2), ... Thus, the routing only relies on the candidate inter-orbit ISLs. The shortest ISL (S5, T5) is then chosen.
		
	 We explain the oscillation in the polar region as follows:

	1) In Fig.\ref{simplified_topology_delay_change}a, the newly reconnecting ISL (S5, T5) is chosen by the routing from S1 to T6.
	
	2) After 85s, (S3, T3) becomes smaller than (S5, T5) due to the satellite moving, shown in Fig.\ref{simplified_topology_delay_change}b. Then the routing will switch to (S3, T3) from (S5, T5).
	
	3) Unfortunately, (S3, T3) is a nearly disconnecting ISL. When the time is 162s, this ISL will be dropped (shown in Fig.\ref{simplified_topology_delay_change}c). The routing thus switches back to (S5, T5).
	
	4) In Fig.\ref{simplified_topology_delay_change}d, (S2, T2) becomes smaller than (S5, T5) at 359s. The routing naturally switches to (S2, T2).
	
	5)  In Fig.\ref{simplified_topology_delay_change}e, (S4, T4) reconnects at 548s. Since the newly reconnecting ISL (S4, T4) has the smallest metric, the routing thus switches to (S4, T4). Meanwhile, a new period will begin.

	From the above analysis, we know that the selected routing is frequently changed, about once per 137s. For many users adopting this routing, they will suffer from the very poor performance, such as delay jitter, loss packets.
	
	Unfortunately, the oscillation will always exist in the polar region. Because all satellites simultaneously move in their own orbits. When a nearly disconnecting inter-orbit ISL disconnects, a new ISL will become the nearly disconnecting ISL. And a newly reconnecting inter-orbit ISL will be created once an inter-orbit reconnects. Along with the link disconnection/reconnection, the metrics of dynamic ISLs continuously vary. The polar oscillation continuously occurs, and meanwhile, those ISLs where polar oscillation occurs constantly change.
	
	Extremely, because the paths between two satellites at the both sides of seam have to go through the polar region, there are about 1260 paths passing the newly reconnecting inter-orbit ISLs and 630 paths passing the nearly disconnecting ones in a polar region. They respectively account for about 30\% and about 15\% of the total paths (4356). Since the satellite constellation has two polar regions, the number of paths and times of switches will be double! 
	
	That means so many paths frequently vary along with the variations of metric and network connectivities, causing polar oscillations and putting great pressure on the network stability. Especially, when many paths switch forth and back on some overhead ISLs, the network performance will suffer from catastrophic damage, such as severe packet loss.

\textbf{Frequent Routing Updates}:
In order to analyze the impact of frequent routing updates, the concept of snapshot based on the network connectivity is applied to model the dynamic topology.  With this concept, a new snapshot is created when some ISLs reconnect or disconnect. And when a new snapshot is created, the corresponding routing table will be updated.	
From the experiment, we can get that the routing updates occur once per 136.8s. In each update, there are about 1050 $\sim$ 1800 end-to-end paths will be forced to change. That account for about $25\% \sim 41\%$ of the total paths (the total is 66 $\cdot$ 66 = 4356). So many changed paths in each routing update will push network into a chaotic state. 

Without consideration of link metric variations, that method can achieve the minimum of routing updates. However, routing updates will fall behind the topology changes, and thus many paths will become suboptimal. Once a routing update is triggered, so many paths will change together, at most $41\%$ of total paths being changed. Hence, the trade-off between the timeliness of keeping consistent with the topology and the cost of routing updates is needed.
%

%
%

	\section{Mitigating Polar Routing Oscillation }
\label{Routing_for_a_Snapshot}
	Designing a stable routing scheme for polar routing oscillation is challenging because of its time-varying property. Not only the oscillation overly frequently occurs, but also the overloaded ISLs where oscillation occurs constantly changes. By take use of the time-varying pattern of oscillation, the routing scheme is designed as follows:

	1)  A link lifetime is defined to locate the ISLs where polar routing oscillation occurs.
	
	2) The lifetime of dynamic ISLs is added into their metrics for eliminating the relative variations of their metrics and meanwhile weaken the influence of ISLs drop/reconnection.
	
	3)  Then the current routing algorithm--Dijkstra algorithm is applied to decrease the number of paths sharing the overloaded ISLs and to avoid the corresponding paths oscillating on those ISLs by taking use of its greedy property.

\subsection{Link Metric}
First, we define a link lifetime for every inter-orbit ISL. The link lifetime varies as the satellite moves. 
The newly reconnecting inter-orbit ISL has the largest lifetime, while the nearly disconnecting inter-orbit ISL has a smallest lifetime. However, when the ISL moves into the polar region, its lifetime becomes zero.
		
When a link reconnects, its lifetime will vary from zero to the largest. While the link disconnect, its lifetime will vary from the smallest to zero. Meanwhile, the variation of link metrics can be mapped to the variation of link lifetime due to the fact that all satellites move at a certain angular velocity. Hence, the variations of inter-orbit ISLs connectivity and metrics can be formulated by the link lifetime. According to the cause of polar oscillation, the ISLs with the largest/smallest lifetime also are the ISL where polar oscillation occurs. 

Then the link lifetime is incorporated into the link metric function, in order to eliminate the relative variations of metrics and weaken the influence of ISLs disconnection/reconnection.
	\begin{equation}	{
	MF=\left \{		 	
	\begin{aligned}
	&\zeta  \quad \text{if and only if the ISL with the largest/smallest}\\  		  
	&\ \quad\text{ lifetime} \\
	&propation\_delay\quad \text{otherwise}
	\end{aligned}
	\right.
	\label{metirc_function}  }
	\end{equation}
 
	For the inter-orbit ISLs with the largest/smallest lifetime, their metrics are set to a constant value $\zeta$. For other ISLs, their metrics are computed by their own propagation delays. The metric $\zeta$ is set to a value that is slightly smaller than the propagation delay of inter-orbit ISL when it just becomes the nearly disconnecting inter-orbit ISL. And the value of $\zeta$ is also smaller than the delay of intra-orbit ISL. 

\begin{figure}[!t]
		\centering
		\includegraphics[height=0.52\linewidth]{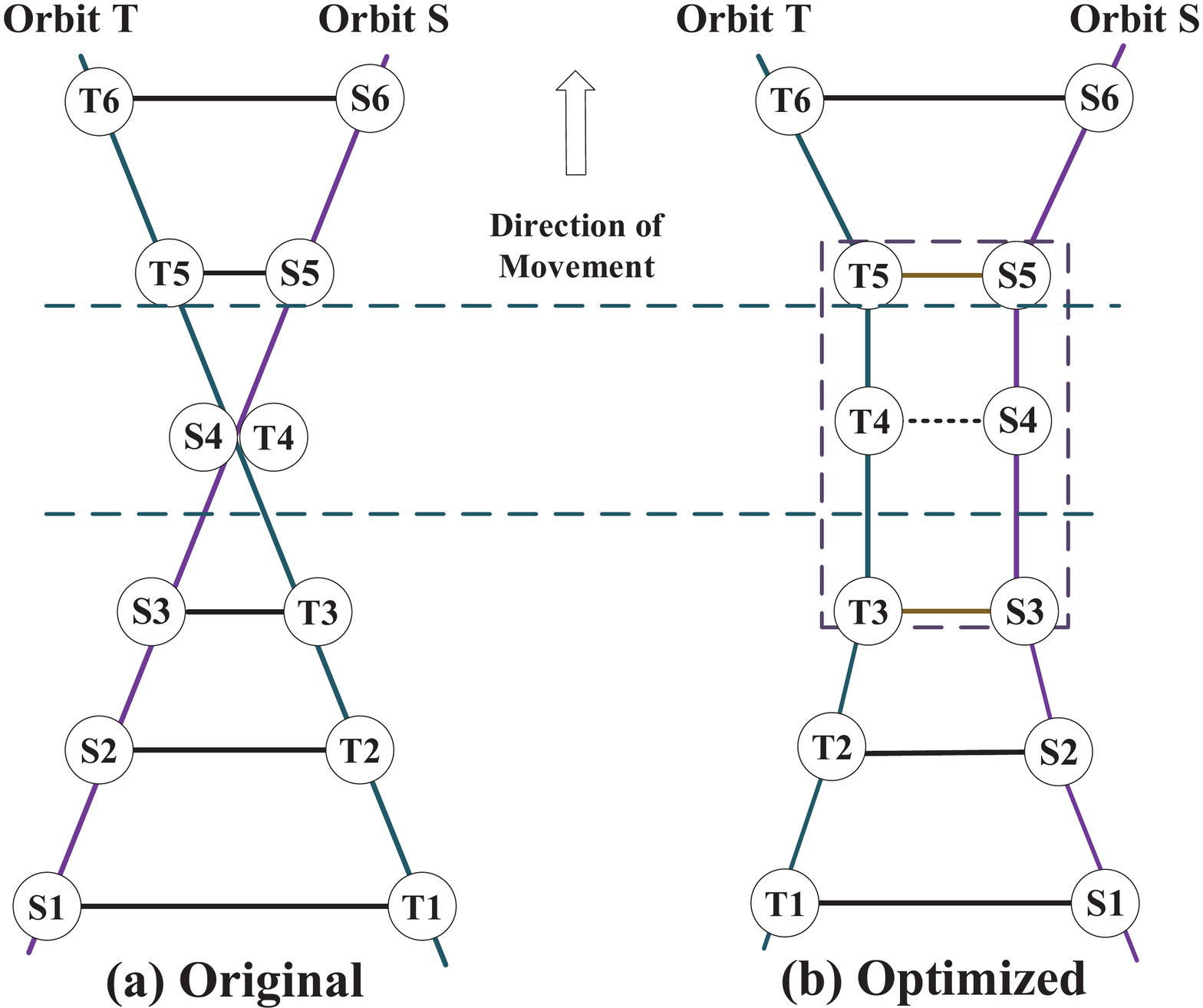}%
		\caption{Comparison between original and optimized topology.}
		\label{link_metric}	
	\end{figure}

    Fig.\ref{link_metric} shows the relative variations of dynamic ISLs are eliminated. Metrics of dynamic links (T5, S5) and (T3, S3) are the same rather than relative variations. Meanwhile, other part of the network topology remains relatively unchanged because metric of (T3, S3) is still smaller than that of (T2, S2), and metric of (T5, S5) is still smaller than that of (T6, S6). Thus, the link function only works for the shortest path passing through the link (T5, S5) and (T3, S3), and impose little influence on others. 

	With this design, the influence of ISLs drop/reconnect is weakened by taking use of movement of the satellite. As shown in Fig.\ref{link_metric}, when (S4, T4) reconnects, it will immediately take the role of (S5, T5) and becomes the link with the largest lifetime; When (T3, S3) disconnects, the (T2, S2) will take its place at once. Meanwhile, the metrics of those ISLs with largest/smallest lifetime remain the same.

	\subsection{Routing Algorithm}
	Based on the link metric function in equation (\ref{metirc_function}) and the topology characteristics of SBI, the Dijkstra algorithm is used to decrease the number of paths sharing the overloaded ISLs and to avoid the corresponding paths oscillating on those ISLs.

\textbf{Example.}
	The Dijkstra is applied to compute the shortest paths from satellite S1 to T6 and from S6 to T1. In Fig.\ref{link_metric}a, with the original topology, both two pairs will go across (S5, T5), because the metric of (S5, T5) is smaller than that of (S3, T3). After a short time, all of them will go across (S3, T3), because the metric of (S3, T3) become smaller than that of (S5, T5). However, in Fig.\ref{link_metric}b with optimized topology, the best path from S1 to T6 will go across (S3, T3), while the best path from S6 to T1 will go across (S5, T5).

\textbf{Proof.}	
	 In Fig.\ref{link_metric}b, the best path from S1 to T6 can be represented by the best path from S3 to T5, according to the optimality principle of Dijkstra algorithm \cite{kleinberg2006algorithm}. When we pick out any two points (satellites) from a certain path, the section between these two points is a part of the original path.
	
	Suppose we pick out the two points -- S3, T5 that has two equal-cost paths. The best path from S3 to T5 is the path passing through (S3, T3) rather than (S5, T5). It can be proved by the greedy poverty of Dijkstra:
	
	1) The cost from point S3 to T5 is constant whether the best path passes through point S5 or T4. As the metric of ISL (T4, T5) is smaller than that of (S5, T5), the cost from S3 to T4 is smaller than that to S5.
	
	2)  According to the greedy property of Dijkstra \cite{kleinberg2006algorithm}, the algorithm will firstly select the point T4 to compute the cost of T5, then the minimal cost of T5 will be achieved.
	
	3) In the next step, the algorithm will select S5 to compute the cost of T5. In this case, the cost of the path passing through S5 is not better than but equal to the current cost of T5. The algorithm will not update the cost of T5, and the father point of T5 will also remain unchanged. Thus, the best path from S3 to T5 still passes through (S3, T3) and point T4.

The path from S1 to T6  will also go across (S3, T3). While the best path from S6 to T1 is across (S5, T5). Their best paths will not be changed until (S3, T3) disconnects or (S5, T5) moves out of this region. Compared with the routing of original topology, the number of paths sharing the inter-orbit ISL (S3, T3) or (S5, T5) is decreased, and corresponding paths switching on those ISLs is avoided.
	 
	 \subsection{Discussion}
	 
	In fact, the routing algorithm is not sensitive to the value of $\zeta$. For keeping the topology of overall network relatively unchanged, it should be set to a value that is slightly smaller than the delay of inter-orbit ISL when it becomes the nearly inter-orbit ISL.
	 
	Besides, other shortest path algorithms may not work in our scheme, such as the Floyd-Warshall algorithm. That is because this algorithm selects the best path from some equal-cost paths by satellite-ID \cite{kleinberg2006algorithm}. For example, let the satellite-ID as follows: T3 $<$ S3  $<$ T4  $<$ S4  $<$ T5  $<$ T6, the routing from T3 to S5 and from T5 to S3 will simultaneously go through (T3, S3). Those inter-orbit ISLs will still suffer from the overloaded routing. When the connectivities of those ISLs change, many paths will be changed and the network performance will be damaged. Moreover, the satellite-ID is constantly varying at the polar region due to the satellite moving. The routing will thus suffer from the instability of satellite ID unless we design a stable numbering scheme for satellites, which is not our concern in this paper.

	\section{Limiting Routing Updates}
\label{Topology_Model}
In order to implement the routing scheme into the dynamic network and meanwhile decrease routing updates, we build a topology model based on the divide-and-merge methodology to partition such topology into a series of snapshots. 

%
%
%
\subsection{Divide Phase}
First, we divide the system period of the dynamic network into a series of time intervals (snapshots). Each snapshot has a start time and a finish time. We use the start time to identify the snapshot, and the finish time of this snapshot is also the start time of the next snapshot. The topology during a snapshot is considered to be fixed.

Let the set $T$ denote all of the divided snapshots that have not been merged. The elements of set $T$ are denoted as $t_1,t_2,\dots,t_n$, with $t_1< t_2 < ... < t_n$. $n$ is the total number of divided snapshots. $t_{k}$ is the start time of snapshot $t_k$, which is also the finish time of its previous snapshot $t_{k-1}$.

In the snapshot $t_k$, the topology is considered as fixed. Thus, the duration $t_{k+1}-t_k$ of snapshot $t_k$ decides the accuracy of the snapshot $t_k$ to reflect the physical topology between the time point $t_k$ and $t_{k+1}$. A smaller $t_{k+1}-t_k$ can accurately reflect the topology variation, including the network connectivity variation and the link metric variation, while a larger $t_{k+1}-t_k$ can induce much topology detail loss.

For correctly reflecting variations of physical topology, $t_{k+1}-t_k$ is set to a very small value. During a short time $t_{k+1}-t_k$, the topology remains relatively unchanged. The continuously dynamic topology then is divided into a series of static topologies, which can reflect variations of physical topology correctly. 

\subsection{Merge Phase}
In this section, our objective is to merge those contiguous snapshots whose topologies are unchanged or only slightly changed into a merged snapshot. When a merged snapshot is created, a new static topology and routing update will be incurred. The routing will be updated at the start time of the first snapshot, and the topology of the first snapshot will be used. In other words,  we use a new merged snapshot instead of those divided snapshots to model the continuously dynamic topology.

Compared with those divided snapshots, the larger duration of merged snapshot may increase the error of reflecting the network dynamics. While the cost brought by routing update will be decreased due to reduced number of snapshots. 
If we merge those contiguous snapshots whose topologies are unchanged or only slightly, both error and cost will remain almost unchanged. 
Thus, the goal of reflecting the dynamics correctly with least number of static topologies can be achieved by minimizing error and cost. 


\textbf{Formulating model.}
From the discussion above, we have a set of divided snapshots $T$ containing $t_1,t_2,..,t_n$. Its subset is denoted as $T_{i, j}$ with $ \{ t_i, t_{i+1},... t_j \} $ indicating $i \le j$.  Snapshots of each subset can be merged into a merged snapshot.
In this case, the goal of that model is to partition the total divided snapshots $T$ into a series of subsets with the minimum penalty. 
The penalty is defined as a weighted sum of:

\begin{itemize}
	\item   $e_{i,j}$: the degree of merged snapshot $T_{i,j}$ failing to reflect the topology variation.
	\item   $w_c$ : the number of merged snapshots into which we partition $T$, times a fixed, given weight $w_c$.
\end{itemize}


\textbf{Definition of error $e_{i,j}$.}
We compute an error $ e_{i,j} $ with respect to each subset $T_{i,j}$, according to the degree of $T_{i,j}$ failing to reflect the topology dynamics. The network topology dynamics between the time point $t_i$ and $t_j$ mainly consists of the link metric variation, and network connectivity variation. The network connectivity variation also includes some ISLs disconnection and reconnection. If these variations occur during a snapshot and are undetected by the model, they will cause errors. The total error $e_{i,j}$ are defined as the sum of errors brought by ISLs reconnection, ISLs disconnection and ISL metric variation, respectively denoted as $e_{i,j}^r$, $e_{i,j}^d$ and $e_{i,j}^m$. And the $w_n$ and $w_m$ are the weights corresponding to the network connectivity variation and link metric variation.
\begin{equation}
e_{i,j}= w_n  \cdot ( e_{i,j}^r + e_{i,j}^d) \ + w_m  \cdot e_{i,j}^m
\end{equation}

Explicitly, if the ISL reconnection during a snapshot is not considered into its topology, the link resource will be under-utilized; if the ISLs disconnection during a snapshot is considered into its topology, packet loss will be incurred after the ISL breaks down; if the link metrics variation during a snapshot is considered into its topology, some shortest paths will become suboptimal.

In fact, the errors are directly caused by the routing variation. Meanwhile, the dynamics of network topology is also directly reflected in the routing variations.
Thus the routing variation is a bridge between dynamics of network topology and errors. From the view of routing variation, these errors are defined as follows.

\begin{equation}{
	\begin{aligned}
	&e_{i,j}^r =  \sum_{r=1}^{R}l_{\delta(r)}^r \cdotp \frac{t_j-t_{\delta(r)}}{t_j-t_i} \\
	&e_{i,j}^d =  \sum_{d=1}^{D}l_{\delta(d)}^d \cdotp\frac{t_j-t_{\delta(d)}}{t_j-t_i} \\ 
	&e_{i,j}^m =  \sum_{m=1}^{M} \frac{\|p_{\delta(m)}^{m} - p_j^{m} \|}{p_{\delta(m)}^{m}}\\
	\end{aligned}
	\label{error}}
\end{equation}

In equation (\ref{error}), the $ R, D, M$ represent the total number of reconnecting ISLs, disconnecting ISLs and suboptimal paths during the merged snapshot $T_{i,j}$, respectively. The suboptimal paths are mainly caused by link metric variations. The occurrence time of these instantaneous events is denoted as $\delta (\cdot)$, which is an element of the set $T$. The number of paths passing link $r$ and link $d$ at the time $\delta(\cdot)$, are receptively denoted as $l_{\delta(\cdot)}^r$ and $l_{\delta(\cdot)}^d$. The impact of ISL reconnection or disconnection on the snapshot $T_{i,j}$ is indicated as $\frac{t_j-t_{\delta(\cdot)}}{t_j-t_i}$.
And $p_{\delta(m)}^{m}$ indicates the length of suboptimal path $m$ at time $\delta(m)$; $p_j^m$ denotes the length of corresponding optimal path at the start time of snapshot $j$. The relative variation between suboptimal and optimal paths is indicated as $\frac{\|p_{\delta(m)}^{m} - p_j^{m} \|}{p_{\delta(m)}^{m}}$.

%




\subsection{Dynamic Programming Algorithm}


\begin{algorithm}[t]
	\caption{Dynamic Programming Algorithm}
	\begin{algorithmic}[1]
		\Function {DPA} {n}
		\State Array O$[1...n+1]$ 
		\State Set O$[n+1] =0$
		\State Compute all pairs error $e_{i,j}$
		\For {$i=n,n-1,...,2,1$}
		\State $\mathsf{O}[i]$=$\min\limits_{i \le k \le n}(e_{i, k} + w_c	 + \mathsf{O}[k+1])$	
		\EndFor
		\EndFunction
	\end{algorithmic}
	\label{Dynamic_progrmming}
\end{algorithm}

\textbf{Designing Algorithm. }
For above model, there may exist many solutions. Now, we design a dynamic programming algorithm to find an optimal solution with polynomial time.

When we have merged some snapshots $T_{1,i}$, the subproblem is to partition the remaining snapshots $T_{i+1,n}$ with minimum penalty. 

%

Suppose the first subset consisting of snapshots $t_1,t_2,\dots, t_{i}$ is optimum, they can be merged into a snapshot $T_{1,i}$. The optimum subproblem $T_{i+1,n}$ is denoted as $\opt(i+1)$. Then we get optimum value of the first subset.

\begin{equation*}{
	\opt(1)=e_{1,i} + w_c+ \opt(i+1)}
\end{equation*}

Because the first subset $T_{1,i}$ and sub-problems $T_{i+1,n}$ are independent, we can remove the first subset and use the same strategy for the sub-problem $T_{i+1,n}$, denoted as $\opt(i+1)$. Then, we can get $\opt(i+1)$ when we determine a $t_{k}$ to produce a merged snapshot $t_{i+1}, ..., t_{k}$.

\begin{equation*}
\opt(i+1)=\min\limits_{i+1 \le k \le n}(e_{i+1,k}  + w_c + \opt(k+1))
\end{equation*}

The algorithm is shown in \textbf{Algorithm \ref{Dynamic_progrmming}}. The set of merged snapshots can be generated according to the result of that algorithm. The routing updates will be triggered at the start time of merged snapshots and the corresponding topology will be used to compute the routing tables.

\textbf{Analyzing Algorithm. }
The algorithm complexity consists of computing the error and cost, and merging the snapshot.

For a known satellite network, it has a fixed number of satellites and ISLs. Thus, the running time of computing the error and cost of each pair $(i, j)$ is only related to the difference of $j$ and $i$, and the complexity is $O (n)$. Since there are $O (n^2) $ pairs $(i, j)$, the total running time is $O(n^3)$.  

The algorithm has $n$ iterations from $n$ to $1$, and the algorithm takes time of $O (n) $ in each iteration. Thus, if the pre-computed error and cost have been determined, the running time of our algorithm is $O (n^2) $.

		\section{Evaluation}
	\label{Experiment}

		\begin{figure*}[t]
		
			\begin{minipage}{0.3\linewidth}
	\centering
	\includegraphics[width=1\linewidth]{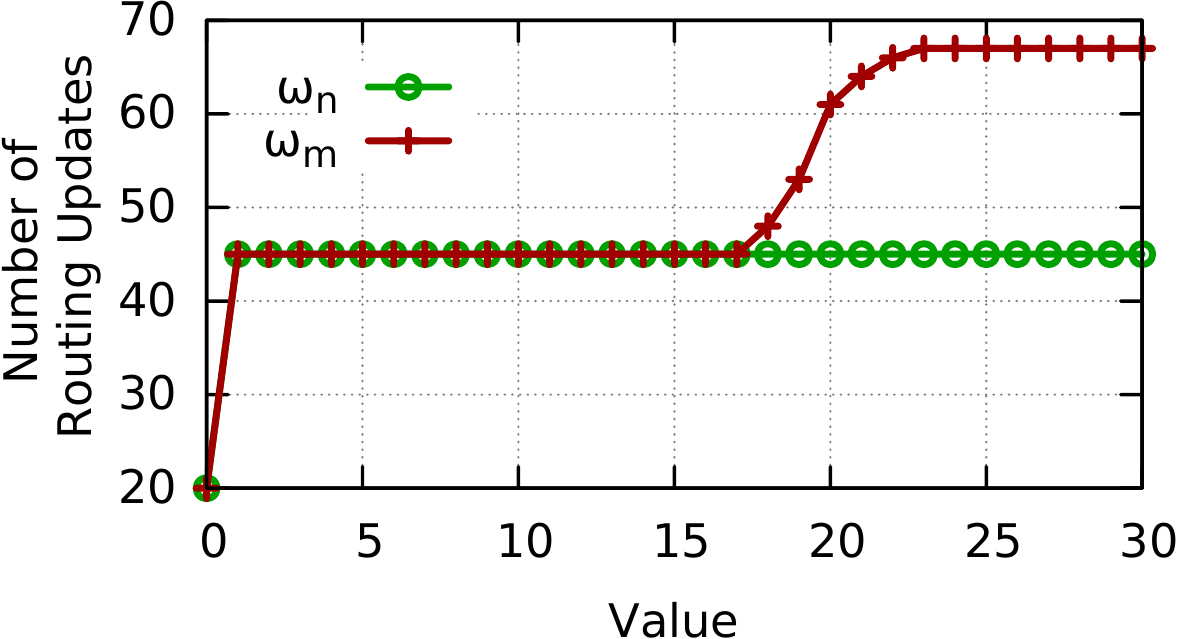}%
	\caption{Effect of weights.}
	\label{update_weight}	
	\end{minipage}
	\hfill
		\begin{minipage}{0.65\linewidth}
			\centering
			\subfloat[Changed paths over a system period] {\includegraphics[width=0.45\linewidth]{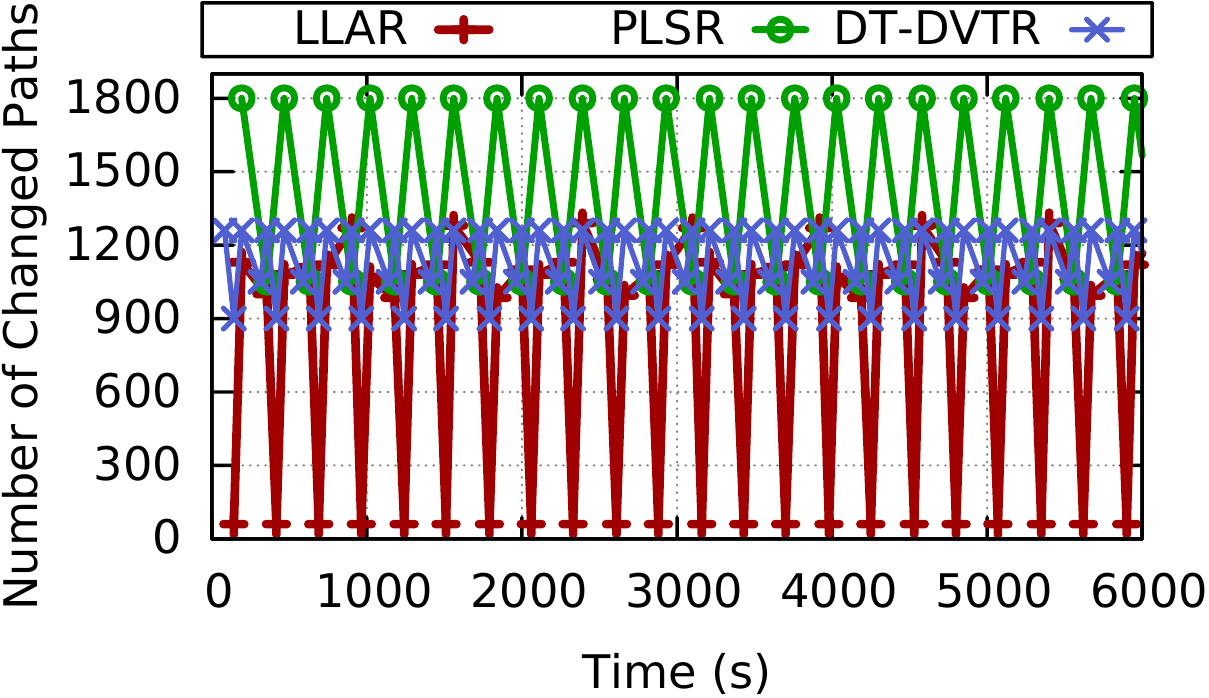}%
				\label{path}}%
			\hfill
			\hfill
			\subfloat[Affected satellites over a system period]{\includegraphics[width=0.45\linewidth]{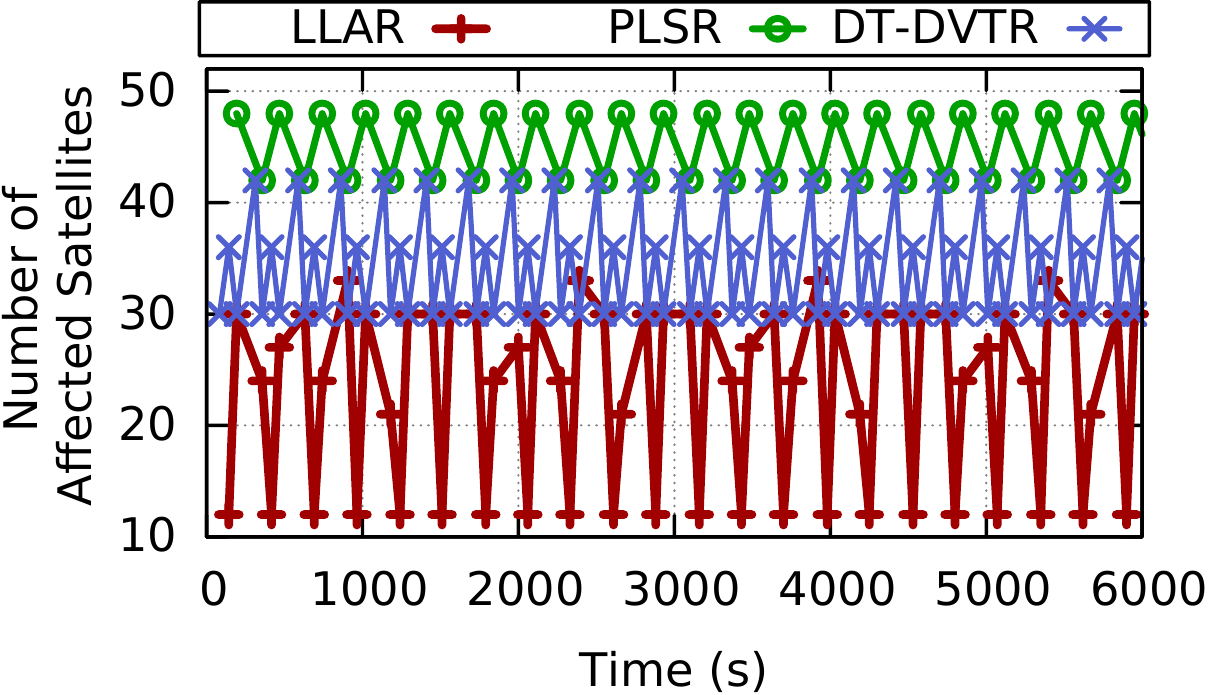}%
				\label{satellite}}
			\caption{Comparison of routing stability among three models.}	
			\label{methods}
		\end{minipage}
		
	\end{figure*}

		\begin{figure*}[t]

		\begin{minipage}{0.28\linewidth}
			\centering
			\includegraphics[width=\linewidth]{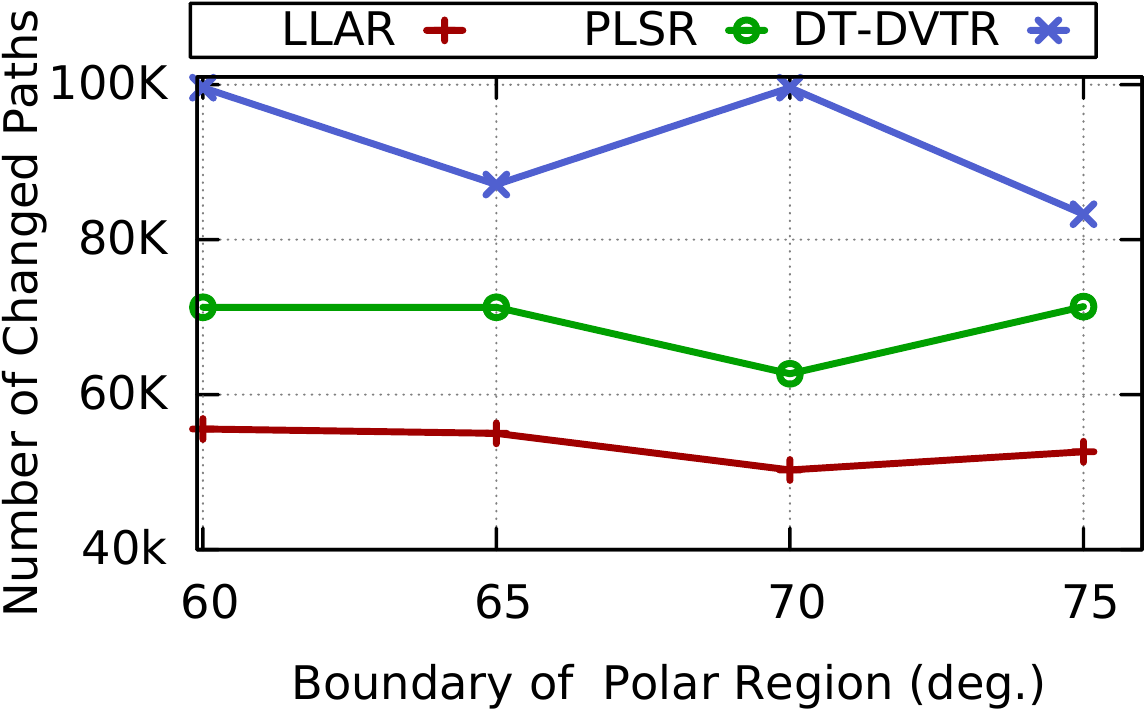}
			\caption{Effect of boundary of polar region}
			\label{changed_boundary}
		\end{minipage}
		\hfill
		\begin{minipage}{0.3\linewidth}
			\centering
			\includegraphics[width=\linewidth]{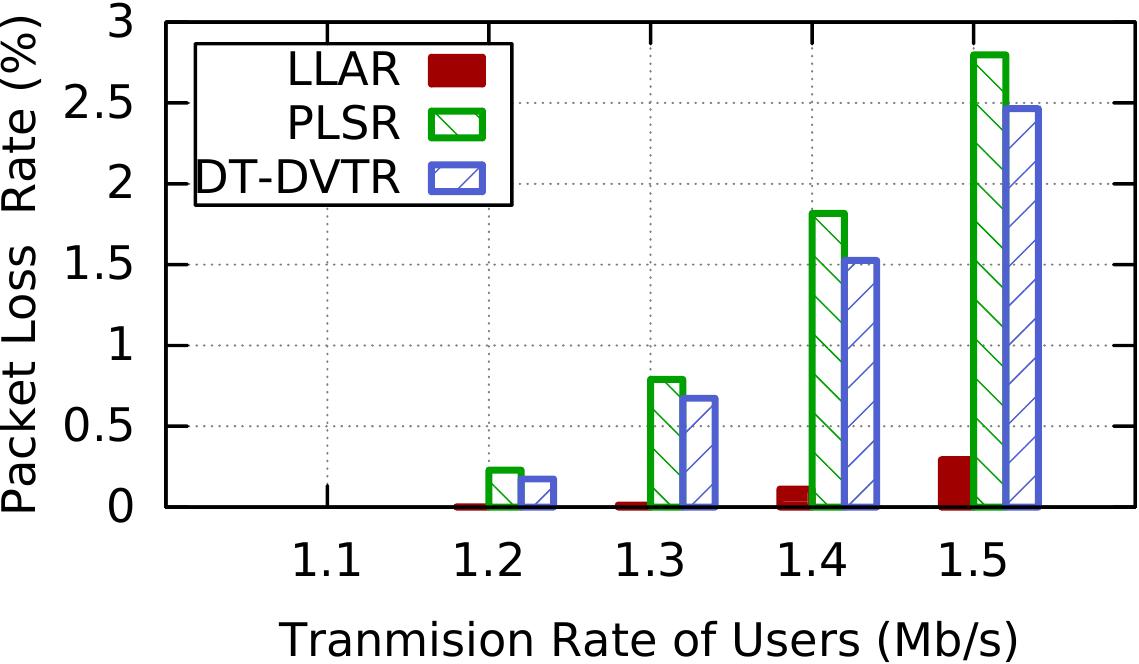}
			\caption{Packet loss rate with load increasing}
			\label{loss_packet}		
		\end{minipage}
		\hfill
		\begin{minipage}{0.29\linewidth}
			\centering
			\includegraphics[width=\linewidth]{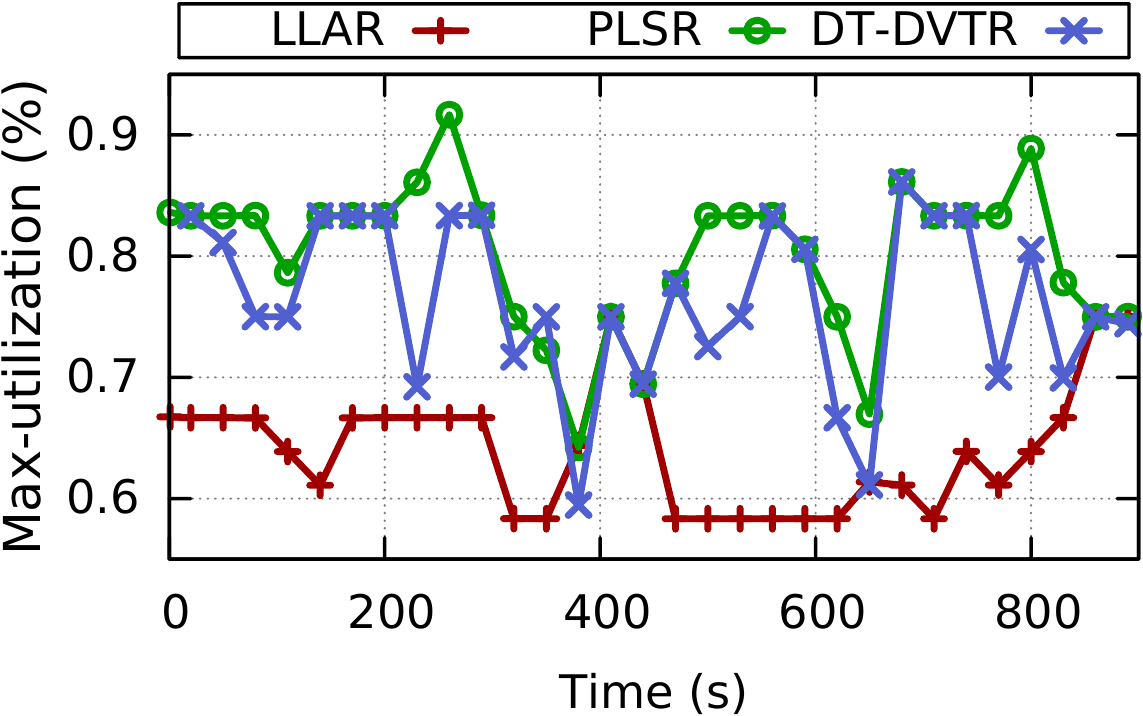}
			\caption{Max-utilization over a simulation time}
			\label{max_utilization}
		\end{minipage}
	\end{figure*}
			
	In this section, we assess the performance of our method in comparison with other two concepts of modeling the network topology, namely DT-DVTR and PLSR. The method of DT-DVTR takes advantage of a fixed time interval to model the dynamic topology, while the PLSR is based on the link drop/reconnect-only. 
	In order to avoid some links unused or broken down between adjacent routing updates in DT-DVTR, we modified its concept that routing update will not only follow its own concept, but also occurs once the network connectivity changes. The interval time of DT-DVTR is set to $60s$ according to \cite{werner1997dynamic}. In our method, the divided snapshot is created like the modified DT-DVTR but the interval time is set to $30s$.  They all apply the Dijkstra algorithm to calculate the routing table at the start time of every snapshot. With the network topology shown in Figure \ref{Iridium_topology} and TABLE \ref{Parameters}, we conduct experiments on NS2.
		
	\subsection{Effect of Parameters}

%

	 Let us analyze the effect of three weights in our model, namely $w_n$, $w_m$, and $w_c$, which respective are the weights of network connectivity variation, link metric variation, and the cost of routing update. First, we set $w_c$ to one, and change the other two. In order to avoid conflicts, we adopt the control variate method. When one of remaining weights is increased step by step, the other is set to zero. When a weight becomes larger, the corresponding indicator becomes more important.
	
	In Fig.\ref{update_weight}, with the weight increasing, the number of routing updates will achieve a steady value. The steady value of weight $w_n$ is equal to the amount of network connectivity changes over a system period. And the steady value of weight $w_m$ is equal to the steady value when we set the two weights to large values. That steady value is also the minimum of snapshots to reflecting the topology dynamics correctly. The minimum is 67,  which is about one-third of total divided snapshots (222). 
	

	\subsection{Analysis of Routing Stability}
		
			
	According to the features of routing stability in SBI, three indicators for routing stability are defined: $(i)$ number of routing updates in a system period; $(ii)$ number of changed paths in each routing update; $(iii)$ number of affected satellites in each routing update. The first two indicate the changes of routing, while the last one indicates the resource overhead for dealing with routing update.

	Fig.\ref{methods} shows that our method can significantly decrease the number of changed paths and affected satellites in comparison with others. The reduced ratios are almost the same, which are about 50\% of PLSR, and about 32\% of DT-DVTR. Our method achieves an average of 761.7 changed paths and 22.7 affected satellites for each routing update. The result of PLSR is worst, which incurs an average of 1425 changed paths and 45 affected satellites in each update. In the worst conditions that do not seldom occur, this method produces 1800 changed paths and 48 affected satellites, which respectively account for 41.3\% and 72\% of the total. The modified DT-DVTR has a slightly smaller number of changed paths and affected satellites than that of PLSR. 
	
	On the other hand, our method requires 67 times of routing updates, which is less than DT-DVTR by 22 and more than PLSR by 24. However, in our method, the total number of changed paths and affected satellites are significantly smaller than others. The reduced ratios of changed paths and affected satellites are about 20\% of PLSR, and about 50\% of DT-DVTR.

	Fig.\ref{changed_boundary} shows that our method achieves the least amount of total changed paths than others no matter how large of the boundary. Compared with PLSR, the total number of changed path in a system period is decreased by $20\% - 26\%$. While that is decreased by $37\%-50\%$ compared with DT-DVTR. 
	
	%

%

	\begin{table}[t]
		\renewcommand{\arraystretch}{1.3}		
		\caption{Simulation parameters}
		\label{Simulation_parameters}
		\centering	
		\begin{tabular}{|c||c|}
			\hline
			Link capacity              & 12 Mb/s \\
			Number of users            & 100 \\ 			
			Packet length              & 1000 B    \\
			Transmission rate of user    & 1.0 -- 1.5Mb/s \\
			Simulation duration        & 900s\\			
			\hline			
		\end{tabular}
	\end{table}

	\subsection{Network Performance }
	In this subsection, the network performance is discussed, and the simulation parameters are shown in TABLE \ref{Simulation_parameters}. As SBI serves all people no matter where they locate and when they access, we pick out 100 users who are randomly distributed in global, the communications between which is random.
	Some important factors are picked out to identify the network performance of different methods, namely, packet loss rate, and maximum link utilization. The utilization of a link is defined as the ratio of the load over its capacity during a certain time. The max-utilization is maximum utilization among all links. From the perspective of max-utilization, utilization near 100\% means that an overloaded link exists in the network and limits the throughput of total network.

	Fig.\ref{loss_packet} shows that our method can drastically decrease packet loss rate regardless of the network load, which is at most 10\% of others. That is because the use of Dijkstra's greedy property in our scene can drastically decrease the number of paths sharing the overloaded ISLs. However, the Dijkstra algorithm in other two methods can't achieve this contribution.  
	
 For fairness, when we analyze the maximum utilization of three methods, the transmission rate of users is set to 1Mb/s where all methods don't lose packets.

   In Fig.\ref{max_utilization}, with the same load, the maximum utilization of our method is the smallest over the experiment. Compared with other two methods, the peak value of the maximum utilization in our method is reduced by about 18\%, and the average maximum utilization is decreased by about 20\%. Meanwhile, the variation of maximum utilization in our method is smoothest.  That means the link in our method has a lightly loaded. Without the loaded links, the network in our method can serve more users can convey more traffic.

		\section{Related work}
	\label{Related_Work}

\textbf{Traditional Routing Schemes} Current routing protocols specific to traditional Internet or ad-hoc network are mostly based on constantly refreshed network state or routing information to get real-time network topologies~\cite{boukerche2011routing,basu2001stability}, which will incur heavy overhead and severely damage network performance for SBI with dynamic topology and limited resources~\cite{alagoz2007exploring,fischer2008topology}.

\textbf{Routing Schemes for Space Network}
In the 90s, many researchers proposed a space-time routing framework aimed to ATM-based space network~\cite{werner1997atm,werner1997dynamic}. However, the success of ground Internet will make it difficult to integrate with these networks. 
With the coming of Satellite-Based Internet, most recent presented routing schemes are based on IP technology~\cite{ ekici2001distributed, henderson2000distributed, akyildiz2002mlsr,fischer2008topology}.
The issue of routing stability in SBI has been focused by some researchers.  However, they mainly considered the convergence issue of traditional routing protocols (e.g., BGP) suffering from topology changes~\cite{berson2009effect, Etefia2010Emulating}. The routing oscillation incurred by route selection has not been considered in their solutions. Besides, they all did not clearly propose the issue of polar routing oscillation.
For avoiding most traffic being drawn into the polar region, \cite {svigelj2004routing} proposed some methods to alternating deflect routing from the polar region for Celestri constellation without network connectivity changes. These methods could achieve load-balance well, but they may aggravate routing instability in term of Iridium-like topology.

\textbf{Topology Models}	
 In \cite{werner1997atm,werner1997dynamic}, the authors adopted a fixed time interval (e.g. 60s) to divide the topology of a system period (simplified DT-DVTR) into a series of static snapshots. This method is very simple, but it may cause route loss/sub-optimality when the link disconnection/reconnection occurs during a fixed time interval. Some authors \cite{gounder1999routing, fischer2008topology} proposed a method of dividing the dynamic topology based on the link connectivity (simplified PLSR). This method can reflect the changes of topology connectivities with least snapshots. However, this method will also cause some details of topology dynamics to be lost, e.g. the delay variation. Moreover, they all have not considered the dynamic characteristics of network topology and the impact of routing updates into their topology models, and thus can not minimize routing updates without damaging the topology dynamics.
Besides, the virtual node (VN) concept has been proposed to hide the dynamic characteristics of network \cite{ekici2001distributed, mauger1997qos}.
However, this approach will push high computational complexity in space devices, and suffers from the outage phenomenon \cite{alagoz2007exploring, henderson2000distributed}. The topology model also may be considered as a special case of DT-DVTR whose the time interval is set as the value of system period divided by number of satellites per orbit.

	\section{Conclusion}
\label{Conclusion}

    In this paper, we deeply analyze the issues of routing stability in satellite-based Internet and provide a novel routing scheme and topology model to solve the issues of polar routing oscillation and frequent routing updates.
Our experimental results show that our methods can make routing more stability and meanwhile achieve a better network performance.     
Besides, our  routing scheme and topology model are simple but effective, which can be the basic element for many routing schemes. The topology model can be easily adapted to other dynamic networks.  
Meanwhile, the routing scheme mainly relies on the current routing technology, which can be a basis for many Internet technologies to be implemented on the space network.
	

{\footnotesize 
\bibliographystyle{IEEEtran}
\bibliography{citations.bib}
}
\end{document}